\documentclass[aps,pre,twocolumn,groupedaddress,showpacs]{revtex4-1}

\usepackage{graphicx}
\usepackage{subfigure}
\usepackage{multirow}

\begin{document}


\title{Heat diffusion in a two-dimensional thermal fuse model}


\author{Glenn T\o r\aa}
 \email{glenn.tora@ntnu.no}
\author{Alex Hansen}%
 \email{Alex.Hansen@ntnu.no}
\affiliation{%
Department of Physics, Norwegian University of Science and Technology, N-7491 Trondheim, Norway\\
}


\date{\today}

\newcommand{\etal}{\textit{et al. }}

\begin{abstract}
We present numerical studies of electrical breakdown in disordered materials using a two-dimensional thermal fuse model with heat diffusion. 
A conducting fuse is heated locally by a Joule heating term. Heat diffuses to neighbouring fuses by a diffusion term. When the temperature reaches a given threshold, the fuse breaks and turns into an insulator.	    
The time dynamics is governed by the time scales related to the two terms, in the presence of quenched disorder in the conductances of the fuses. 
For the two limiting domains, when one time scale is much smaller than the other, we find that the global breakdown time $t_r$ follows $t_r\sim I^2$ and $t_r\sim L^2$, where $I$ is the applied current, and $L$ is the system size. However, such power law does not apply in the intermediate domain where the competition between the two terms produces a subtle behaviour. 
\end{abstract}

\pacs{62.20.M-, 46.50.+a, 07.05.Tp}


\maketitle

\section{Introduction}
Motivated by important aspects such as failure prediction and material improvement, resistor network models \cite{ARandom, kahng88, Annealed, pennetta00} have been intensively used for numerical studies of mechanical and electrical breakdown phenomena in disordered media. The \textit{quasi-static} fuse model \cite{ARandom}, although very simple, reproduces the basic damage regimes in breakdown phenomena \cite{FiberandFuseModel, Regimes}. In the infinite disorder limit \cite{perc} the percolation regime is present. The fracture is totally disorder driven, and fuses burns out randomly until global breakdown is reached. If the disorder is small, there is little precursor damage until a single fracture is developing from one end of the system to the other. This is the nucleation regime. 

However, the quasi-static fuse model contains no real dynamics, and does not capture time dependency in correlations caused by the local currents. Dynamical effects has been included in the fuse model to study the elasticity problem \cite{zap97}. Sornette and Vanneste \cite{vs92,ThermalFuse} developed a model for electrical breakdown and plastic deformation, which they referred to as the dynamic thermal fuse model. The temperature of a fuse was governed by a general Joule heating term, $i^b/g$, and a heat loss term, $-aT$, which can be considered as a simplification of a full spatial diffusion description. 
This model has been experimentally realized by Lamaign\`ere \etal \cite{Experimental}, and later Mukherjee \etal \cite{Predictable,Predictable2} by studying electrical breakdown of carbon-polymer composites. This shows that the thermal fuse model is able to capture some of the phenomenology of breakdown in disordered media, like critical behaviour in the breakdown time. However, the model does not take into account the correlations due to heat diffusion between fuses. Thermal interaction with neighbour fuses has previously been studied by Pennetta \etal \cite{pennetta00,pennetta01} in a biased percolation model.
But this model implies instant thermalization of the fuses, and thereby neglects time dependent effects. 

With the thermal fuse model as a base we introduce spatial heat diffusion in the system. We study how the interplay of quenched disorder, current enhancement effects and heat diffusion give rise to time dependent effects which may seem counterintuitive with respect to the quasi-static fuse model and the biased percolation model.


\section{Model}
\begin{figure}[!htbp]
\resizebox{0.3\textwidth}{!}{\includegraphics{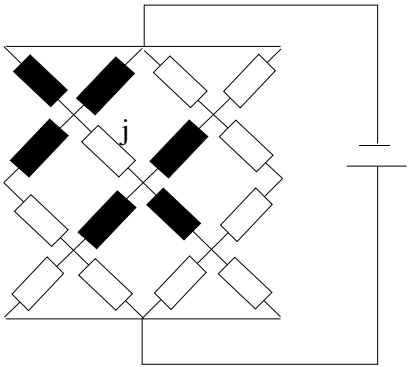}}
\caption{\label{fig:network}A small network of $L=4$. The black fuses are the 6 heat-exchanging neighbours to fuse $j$. 
	}
\end{figure}

Our simulations are based on the thermal fuse model proposed by Sornette \etal \cite{vs92,ThermalFuse}. The model consists of a square lattice oriented at $45^{\circ}$ with respect to the two boundaries opposite of each other, which act as busbars (see Fig.~\ref{fig:network}). Each bond in the lattice is an electric fuse which behaves like an ohmic resistor when intact. A voltage is applied over the busbars which induces a total current $I$ in the lattice. To each fuse $j$ we assign a conductance $g_j$ from a power-law distribution $p(g)\propto g^{-1+\beta}$. The conductances are generated from $g_j=x_j^B$, where $x_j$ is a uniformly distributed random number in the interval (0.5,1.5), and $B=\beta^{-1}$. We call $B$ the disorder parameter. The values of the conductances are set at the beginning and never changed, corresponding to a quenched disorder. 

Our model differs from the thermal fuse model by Sornette \etal in the sense that the term $-aT$ is replaced by a spatial diffusion term. We also use a fixed $b=2$, which corresponds to the Joule heating effect. There is no loss of heat at the boundaries, so the system will always reach global breakdown for $I>0$. 
We then arrive at the following heat equation for our model in two dimensions in the continuum limit.

\begin{equation}
\frac{\partial T}{\partial t}(x,y,t) = \frac{i^2}{gC}(x,y) + D\nabla^2 T(x,y,t), 
\label{3}
\end{equation}

with boundary condition $\partial{T}/\partial n = 0$ on the top and bottom boundary. Periodic boundary conditions are applied in the horizontal direction. The heat capacity is $C=1$ for every fuse, and $D$ is the thermal diffusivity. The evolution in time is given by explicit 
Euler integration, and adaptive timestep is used. When a fuse reaches its temperature threshold $T_r$, chosen equal for all fuses, it breaks and irreversibly turns into an insulator. This is the way the temperature field reacts back on the current field. All the fuses start at equal temperature. When a fuse burns out, the current redistributes itself instantaneously in the network. The network will then heat up until another fuse burns out. The total current is kept constant during the fracture process, with $I=1$ for the results herein, if not specified otherwise. The fuses interact with the 6 nearest neighbours through heat diffusion (see Fig.~\ref{fig:network}). 
Note that if more than one fuse reach the threshold $T_r$ within the same time step, those fuses are broken before the temperature field is updated. This ensures that the currents are instantaneously redistributed. The current distribution is calculated by the conjugate gradient method.

\section{Two competing time scales}
\begin{figure*}[!htbp]
\subfigure[]{
\begin{tabular}{ccc}
		\multicolumn{3}{c}{$B=1$}\\
		\includegraphics[width=0.3\linewidth]{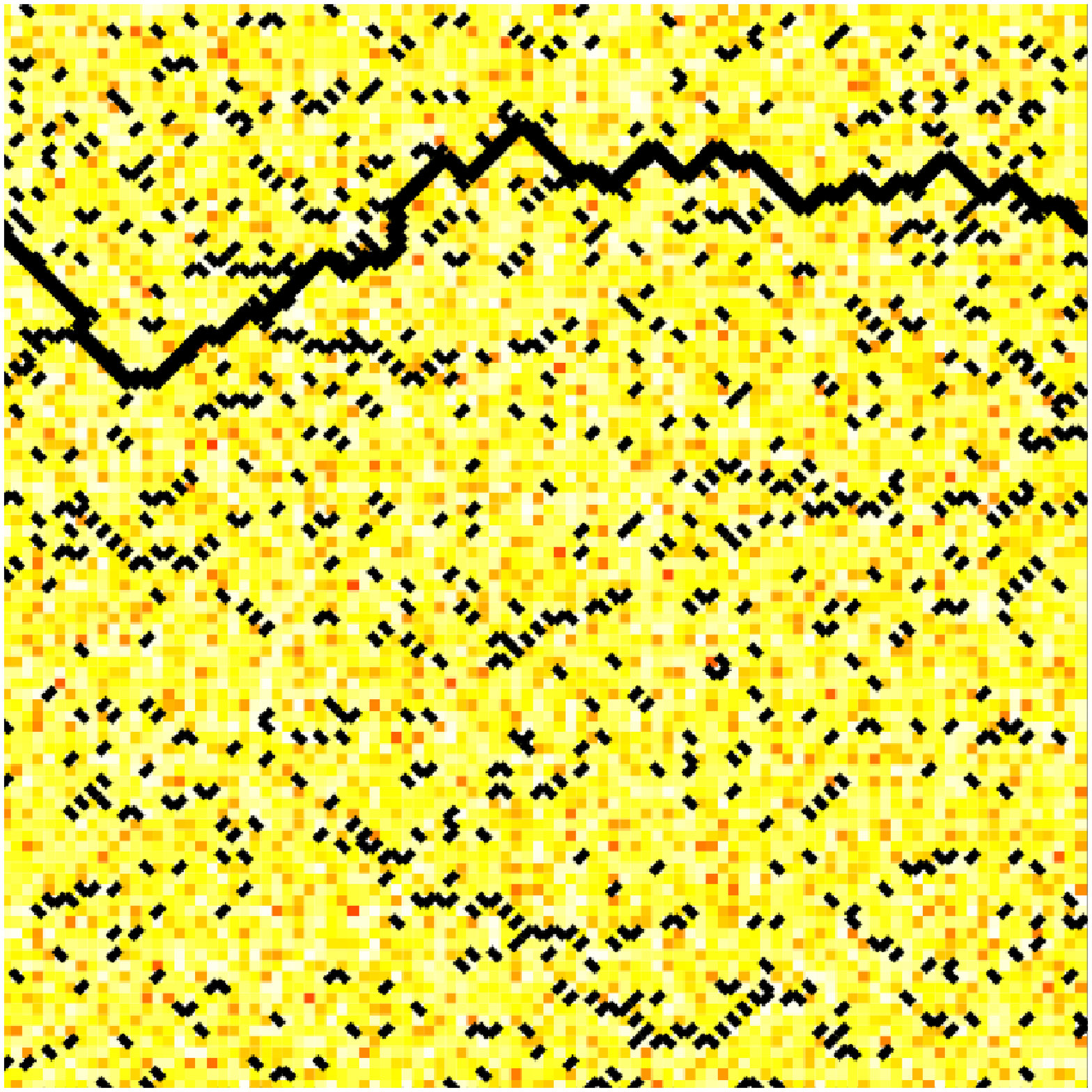}&
		\includegraphics[width=0.3\linewidth]{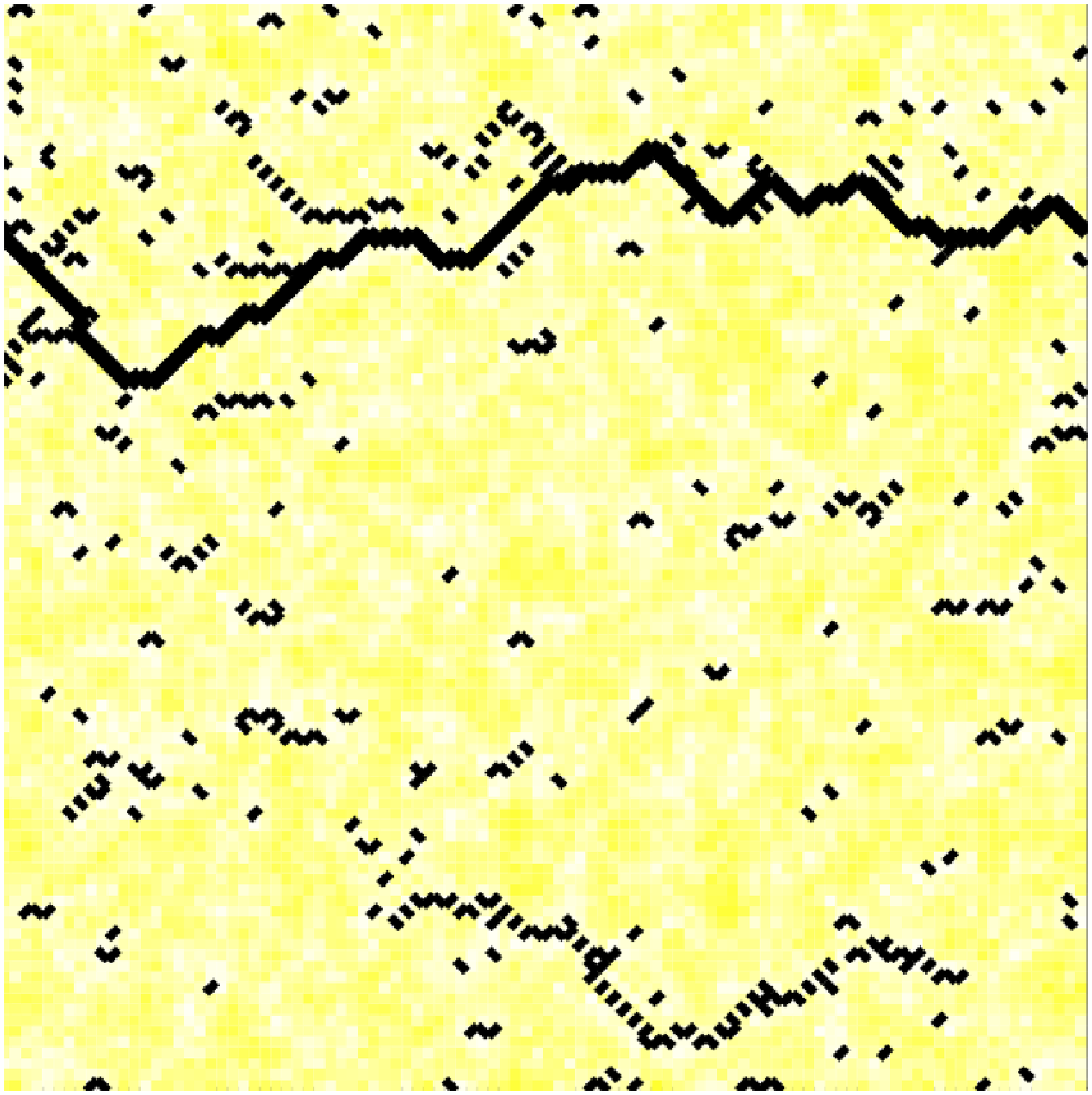}&
		\includegraphics[width=0.3\linewidth]{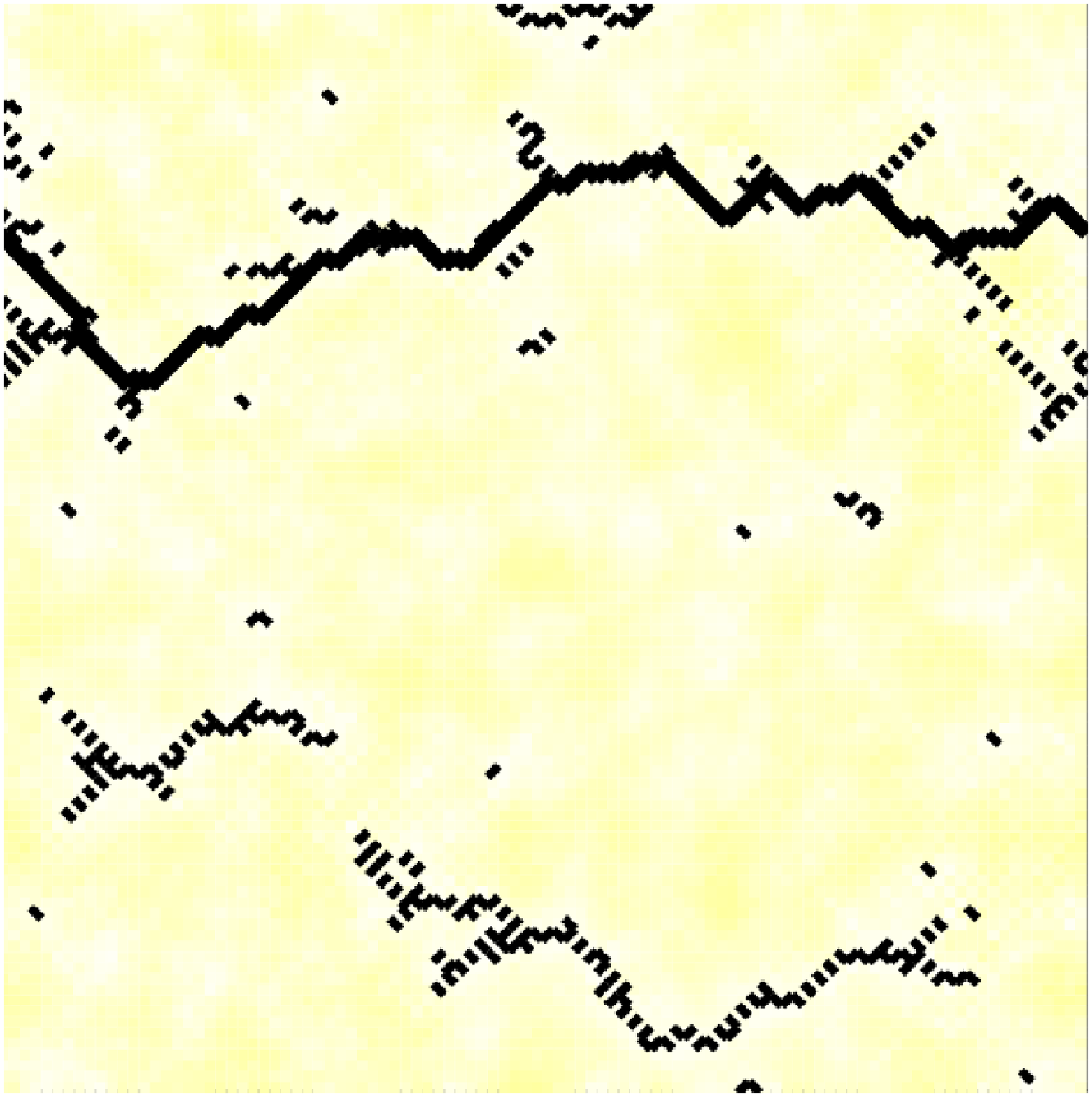}\\ 
		$D=0$  &$D=0.1$  &$D=0.5$ \\ 
\end{tabular}

}\\
\subfigure[]{
\begin{tabular}{ccc}
		\multicolumn{3}{c}{$B=5$}\\
		\includegraphics[width=0.3\linewidth]{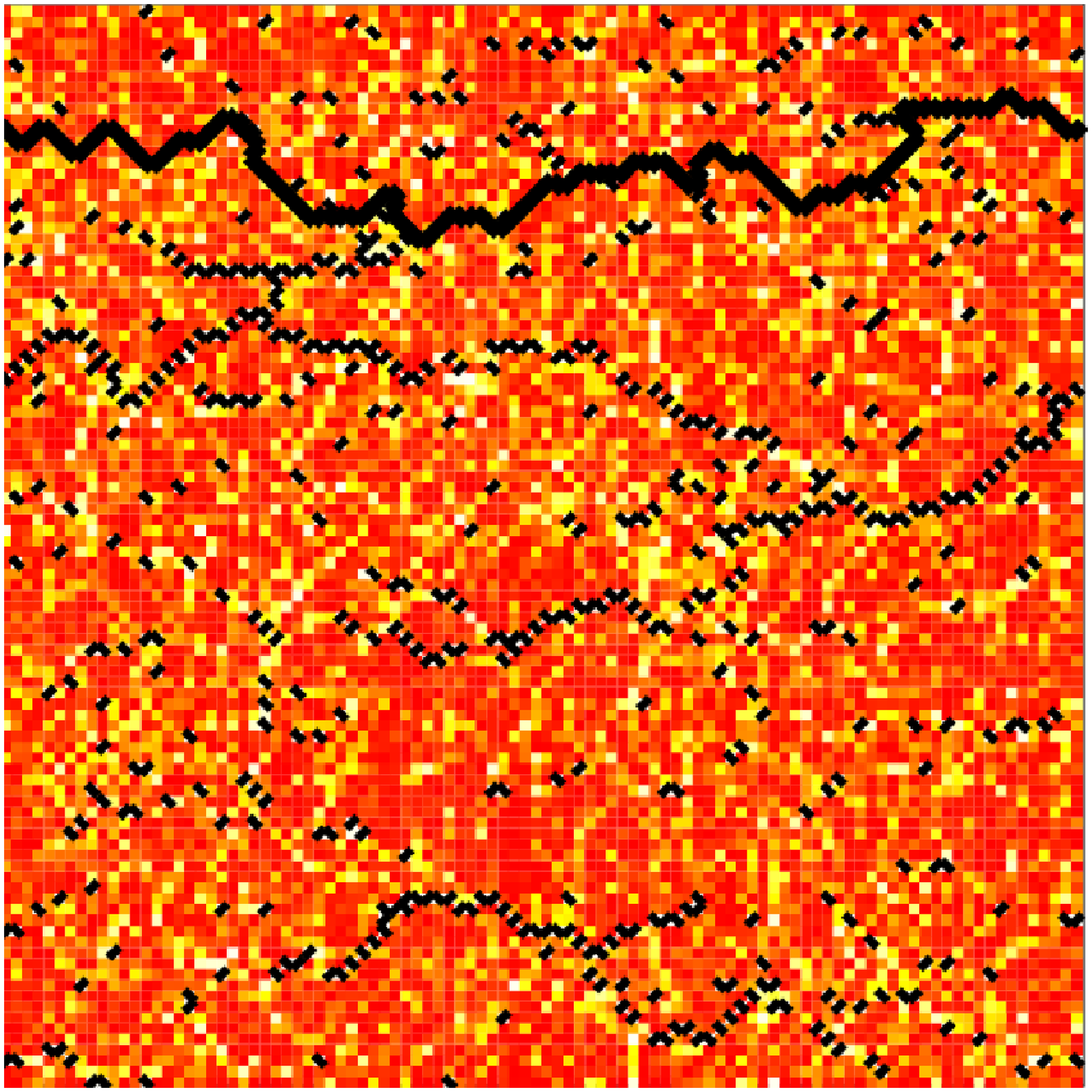}&		
		\includegraphics[width=0.3\linewidth]{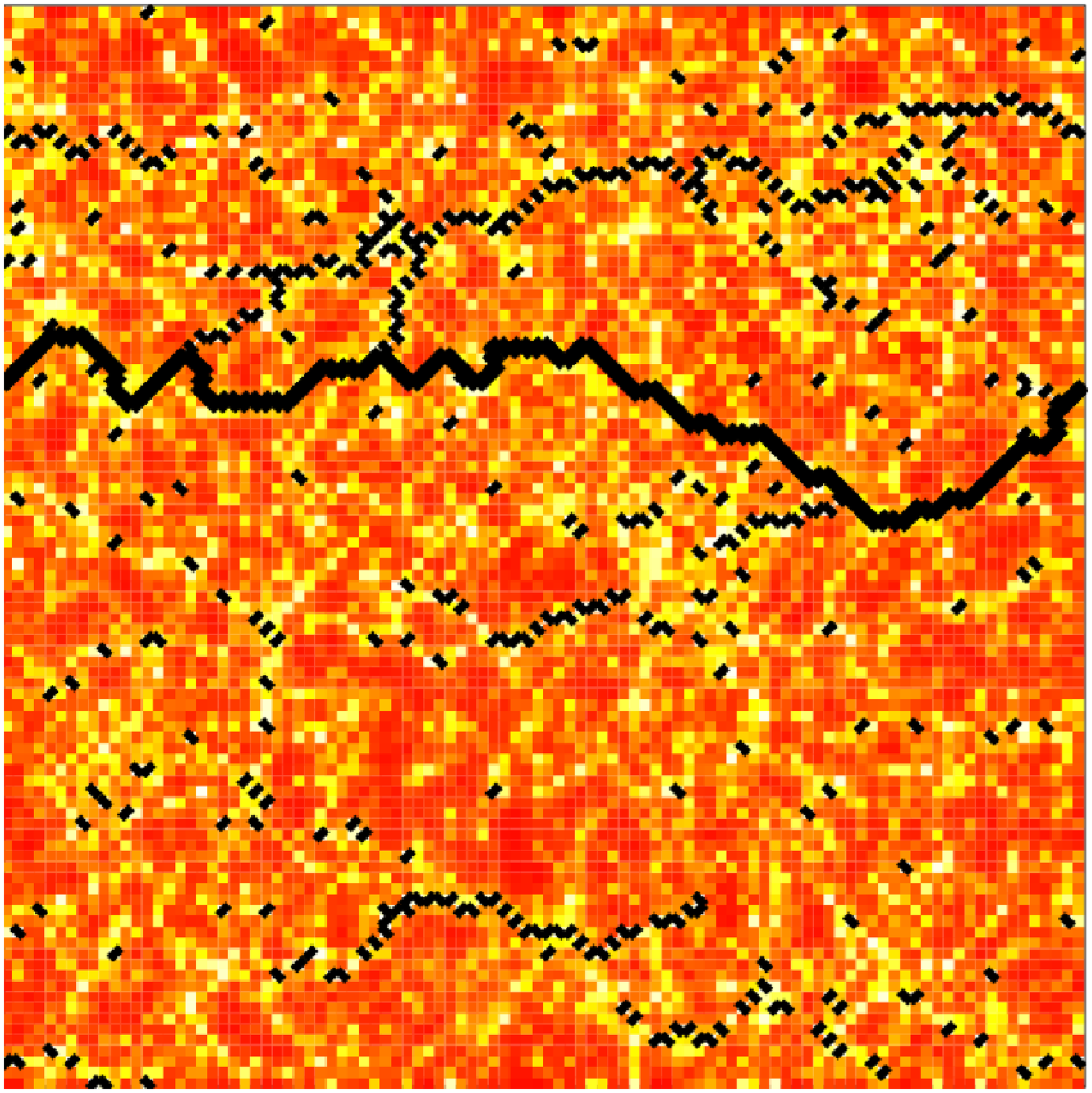}&
		\includegraphics[width=0.3\linewidth]{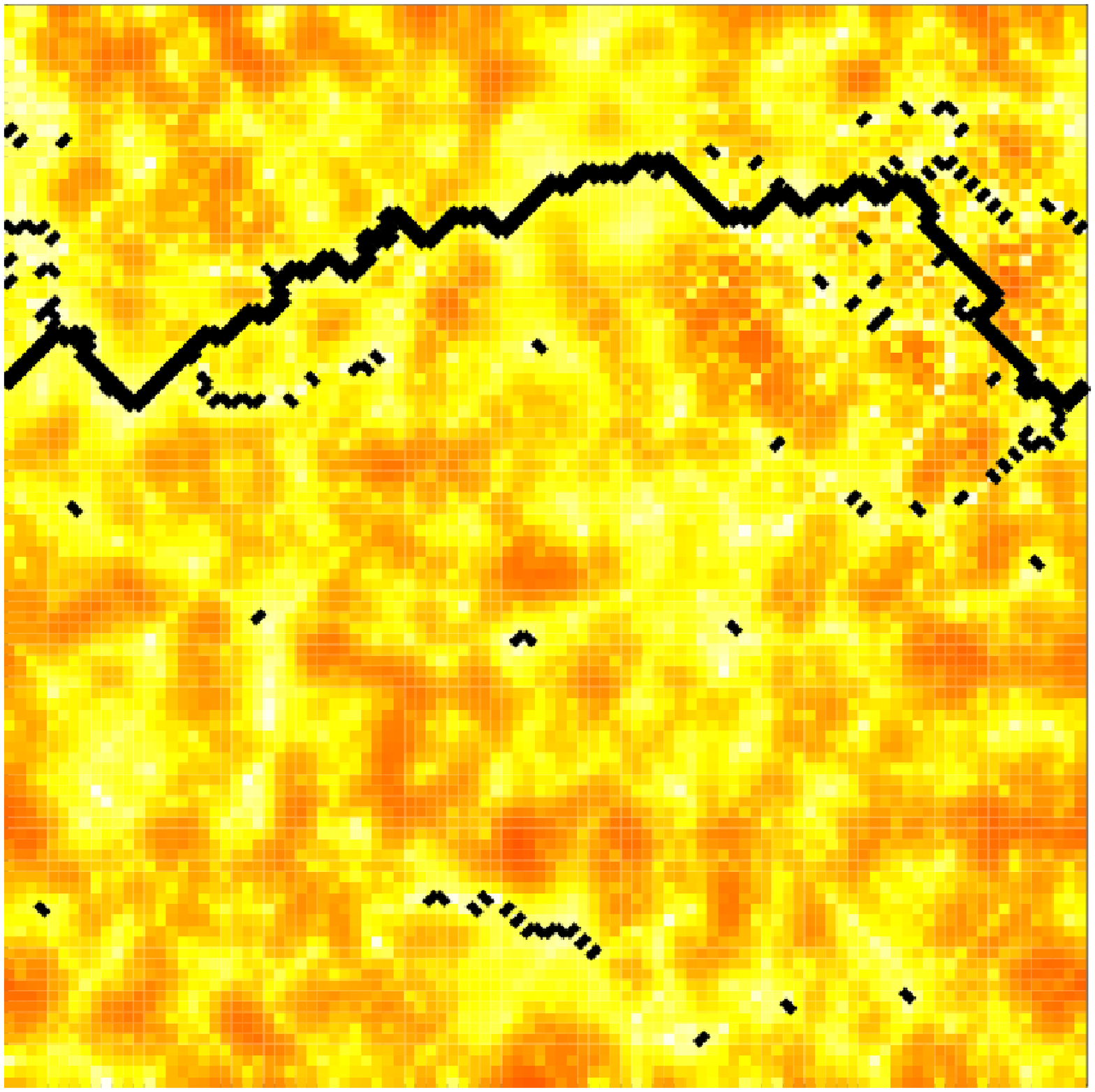}\\					
		$D=0$ &$D=0.1$  &$D=0.5$ \\
		
\end{tabular}
\label{fig:patterns1b}
}

\begin{tabular}{lr}
\multicolumn{2}{c}{
\subfigure{
\includegraphics[width=4.7cm]{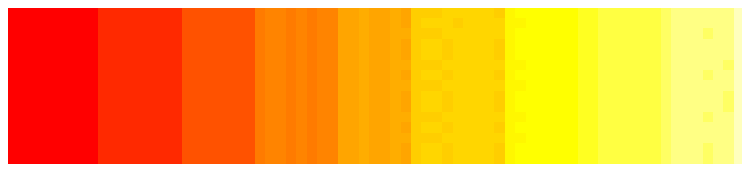}
}}\\
 Cold & Hot
\end{tabular}

\caption{\label{fig:patterns}(Color online) Temperature distribution with fracture patterns at the moment global failure is reached. The fracture which disconnects the network in two pieces is outlined. A system size of $L=100$ is used. (a) $B=1$ and (b) $B=5$. A fixed seed is used for the random number generation so that the effect of the diffusion is clear. 
}
\end{figure*}

A series of final fracture patterns for different values of $D$ and $B$ are shown in Fig.~\ref{fig:patterns}. The fracture which disconnects the network in two pieces is outlined, and the temperature field is indicated by colors. $B=1$ yields a uniform distribution, while $B>1$ gives a broader distribution with a tail towards large conductances. Note the lower cutoff at $0.5^B$, and upper cutoff at $1.5^B$ in the distribution. 

The time dependence is governed by the two time scales $\tau_0=gCT_r/i^{2}$ and $\tau_1=\xi^2/D$. The competition between these time scales generates different domains of behaviour. For $\tau_0<<\tau_1$ the domain is referred to as the large-current domain. If we consider $D=0$, the large-current domain is present for all values of $B$.
By comparing the fracture patterns for $B=1$ and $B=5$ we see that for $B=5$ a few large cracks appear, while for $B=1$ the pattern is more diffuse, with single broken fuses. Due to the broader distribution for $B=5$, the small conductances will heat up much faster than the large conductances, and the cracks will grow along pathways of small conductances. This is contrary to what the quasi-static fuse model gives, where increased disorder results in more single broken fuses, and a diffuse breaking pattern. This difference can be attributed to where the disorder is applied in the two models. In our model the disorder is in the conductances, while in the quasi-static fuse model the disorder is in the thresholds of the fuses. 
 
As $D$ is increased, an intermediate domain appears. Since the diffusion smoothens temperature differences, the disorder in the temperature field decreases, and the effect of the redistribution of the local currents becomes more apparent. This is visible in form of a less diffuse damage pattern, and a more localized crack (see Fig.~\ref{fig:patterns} for $D=0.1$ and $D=0.5$). 
 
Sufficiently large values of $D$ yields $\tau_0>>\tau_1$, and the small-current domain appears. This is dominated by the redistribution of the local currents, and is characterized by a single crack disconnecting the system, with little precursor damage. Similar behaviour appears for low disorder in the quasi-static fuse model.  

\begin{figure}[!htbp]
\includegraphics[width=0.45\textwidth,clip]{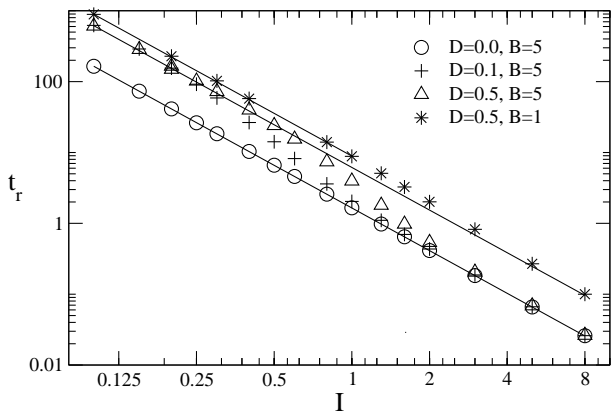}
\caption{\label{fig:I}Log-log plot of the breakdown time $t_r$ as a function of $I$. The solid lines correspond to Eq. \ref{eq:I2}. The data points are averaged over 10 samples for $L=100$. 
}
\end{figure}

The three domains are visible when we plot the breakdown time $t_r$ as a function of $I$ on a log-log scale. This is shown in Fig.~\ref{fig:I}. For both the small- and large-current domain we find a good fit with the power law 

\begin{equation}
t_r\sim I^{-2}.
\label{eq:I2}
\end{equation}

To recover this, it is sufficient to show that the order of local breakdown events is independent of the value of $I$. For the large-current domain we can neglect the diffusion term in Eq. \ref{3}. Then we realize that the first fuse to break will not change when $I$ is changed, and that the local currents are proportional to the applied current, $i\propto I$, between each breaking event. It follows that rest of the breaking sequence will follow in the same order when $I$ is changed. A more complete derivation of Eq. \ref{eq:I2} is given in Ref. \cite{ThermalFuse}.

In the small-current domain the temperature differences are small, and the value of the diffusion term in Eq. \ref{3} will only be significant in a short period of time after each breaking event. Hence, it is reasonable to assume $i\propto I$ between each breaking event, and we obtain Eq. \ref{eq:I2} for this domain as well. 

\begin{figure}[!htbp]
\includegraphics[width=0.45\textwidth,clip]{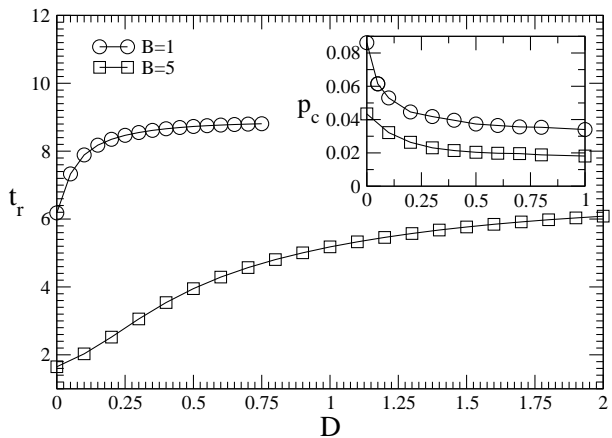}
\caption{\label{fig:ttf}Time to failure $t_r$ as a function of $D$ with $I=1$. Inset: percolation threshold $p_c$ as a function of $D$. The data points are averaged over 10 samples for $L=100$.}
\end{figure}

The presence of the three domains for different values of $I$ is in agreement with the thermal fuse model by Sornette \etal \cite{vs92,ThermalFuse}, and experimental results. Lamaign\`ere \etal \cite{Experimental} fitted the time-to-failure data with Eq. \ref{eq:I2} for large values of $I$, and $t_r\sim (I-I_c)^{-2.1}$ for small values of $I$. Mukherjee \etal \cite{Predictable2} found a good fit with $t_r\sim (I^2/I_c^2-1)^{-1}$. By letting $I_c\to 0$, which is the case in our model due to no heat loss, these fitting laws are consistent with Eq. \ref{eq:I2}.
In the intermediate domain there is a crossover from the small-current to the large-current domain. In this domain the order of local breakdown events depends strongly on the value of $I$, and the time dependence is more subtle. Fig.~\ref{fig:ttf} shows the breakdown time versus $D$. We find that larger values of $D$ result in longer $t_r$, since high temperatures are more effectively smoothened out. This result is qualitatively in agreement with the biased percolation model \cite{pennetta00}. However, a probabilistic approach to the disorder was used, while our model uses quenched disorder in the conductances of the fuses. This difference manifests itself in the impact $D$ has on the percolation threshold $p_c$. This is shown in the inset of Fig.~\ref{fig:ttf}. Diffusion causes less disorder in the temperature field, hence we find that $p_c$ decreases as $D$ increases. This is contrary to what the biased percolation model gives, which approaches ordinary percolation ($p_c=0.5$) for vanishing disorder in the temperature field. 
We also find that the difference in $t_r$ between the two limiting domains for $D=0$ and $D>>0$ in Fig.~\ref{fig:ttf} is smaller for $B=1$ than for $B=5$. This difference reflects the time scale, $\tau_1$, at which temperature differences are smoothened out, and $\tau_1$ decreases with decreasing disorder. 

\section{Scaling analysis}   
Based on analogy to percolation theory \cite{Percolation}, we assume the following finite-size scaling relation

\begin{equation}
\label{eq:x1}
t_r\sim L^{s}.
\end{equation}

\begin{figure}[!htbp]
\includegraphics[width=0.45\textwidth,clip]{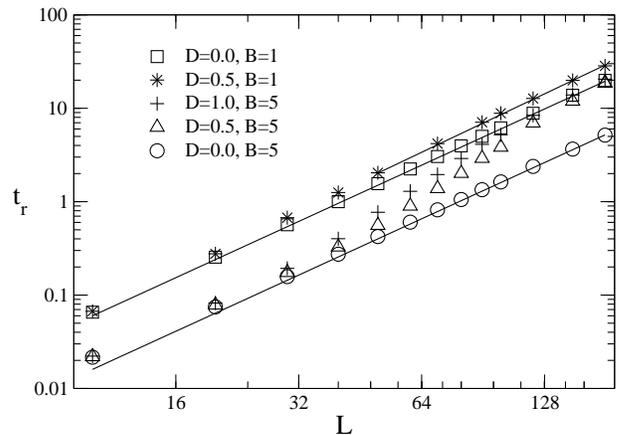}
\caption{\label{fig:tr}Log-log plot of the breakdown time $t_r$ as a function of $L$. The solid lines correspond to Eq. \ref{eq:x1} with $s=2$.}
\end{figure}

Fig.~\ref{fig:tr} shows a log-log plot of $t_r$ as a function of $L$. The total electric resistance $R$ is independent of $L$ in a homogeneous two-dimensional system. We assume this for our model also, and we get $i\propto 1/L$. Since $i\propto I$ in both the small- and large-current domains, it follows that $t_r\sim L^2$. Hence, for small values of $L$, the large-current domain appears ($\tau_0<<\tau_1$). A crossover to the small-current domain ($\tau_0>>\tau_1$) is observed for sufficiently large values of $L$ with $D>0$. 

However, in the intermediate domain $i\propto I$ is not a valid assumption, and $I$ can not be replaced by $1/L$ in the function for $t_r$.

\section{Roughness of the fracture}
We define the width of the fracture surface as the height-height fluctuations $w=(\langle z^2 \rangle-\langle z \rangle^2)^{1/2}$, where $z$ is the height from the bottom of the network to the fuses that belong to the fracture surface, i.e the crack which disconnects the network. Studies of the quasi-static fuse model in two dimensions establish that the width scales as $w\sim L^{\zeta}$, with $\zeta=0.7$ within 10\% accuracy for different threshold distributions \cite{Roughness,bakke08}.                                                                                                   

\begin{figure}[!htbp]
\includegraphics[width=0.45\textwidth,clip]{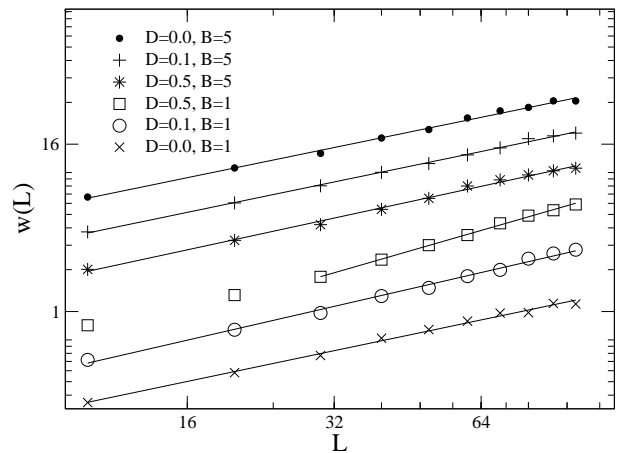}
\caption{\label{fig:roughness}	A log-log plot of $w$ as a function of $L$. The solid lines are linear fits to the data, and gives the roughness exponents shown in Tab. \ref{tab:roughness}. The data for different parameter values are shifted vertically for clarity.
	}
\end{figure}

We generated between 1000 and 100 samples of sizes from $L=10$ to $L=100$, for various values of $B$ and $D$. 
A log-log plot of $w$ as a function of $L$ is shown in Fig.~\ref{fig:roughness}. The slopes of the linear fits give the exponents $\zeta$ listed in Tab. \ref{tab:roughness}. 
\begin{table}[!hbp]
\caption{\label{tab:roughness}Values of $\zeta$ for different thermal diffusivities and disorder parameters.}
\begin{ruledtabular}
\begin{tabular}{ccc}
$D$ & $B$ & $\zeta$\\
\hline 
$0.0$ & $5$ & $0.72\pm 0.1$\\ 
$0.1$ & $5$ & $0.72\pm 0.1$\\ 
$0.5$ & $5$ & $0.76\pm 0.1$\\ 
$10$ & $5$ & $0.77\pm 0.1$\\
$50$ & $5$ & $0.76\pm 0.1$\\
$0.0$ & $1$ & $0.74\pm 0.1$\\
$0.1$ & $1$ & $0.80\pm 0.1$\\
$0.5$ & $1$ & $1.0\pm 0.1$

\end{tabular}
\end{ruledtabular}

\end{table}
The roughness exponent $\zeta$ seems to be independent of which time scale is dominating for $B=5$, i.e large disorder. We obtain $\zeta=0.75\pm 0.1$, which is in the range of the global roughness exponent observed in the quasi-static fuse model. For $B=1$ the roughness exponent is highly dependent on $D$, and is approaching unity for increasing $D$. It means that the fracture is guided by the structure of the network, and this is a trivial regime of behaviour.

\section{Divergence of the resistance}
\begin{figure}[!htbp]
\includegraphics[width=0.45\textwidth,clip]{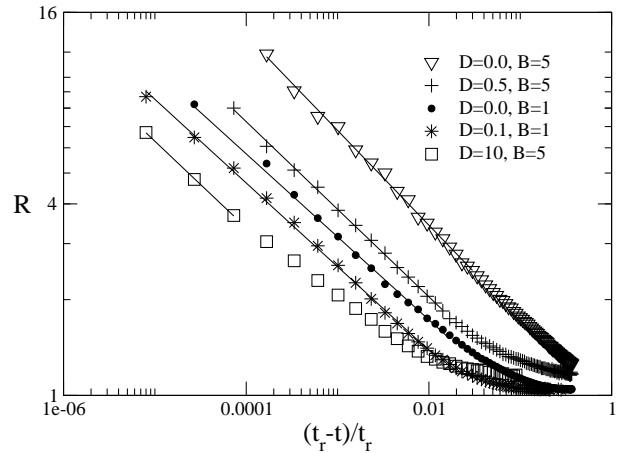}
\caption{\label{fig:rt}Log-log plot of $R$ as a function of $(t_r-t)/t_r$ for different values of $B$ and $D$. The data points are averaged over 50 samples of $L=100$. The solid lines have slopes between 0.26 and 0.30, giving $\alpha \approx 0.28$. For $D=10$ we see that the critical region is very small, close to $t_r$.
	}	 	
\end{figure}

In the study by Lamaign\`ere \etal \cite{Experimental} they found that the total electrical resistance $R$ follows the power law 

\begin{equation}
R\sim (t_r-t)^{-\alpha},
\label{eq:alpha}
\end{equation}

in a critical region close to $t_r$. A value of $\alpha_{\text{3D}} \approx 0.65$ was obtained. A log-log plot of $R$ versus $(t_r-t)/t_r$ for our model is shown in Fig.~\ref{fig:rt}. The obtained value is $\alpha = 0.28\pm 0.05$. This is in agreement with the reported value for the thermal fuse model \cite{ThermalFuse}, and seems to be independent of the thermal diffusivity and disorder parameter. However, the critical region for power law scaling moves closer to $t_r$ as the small-current region is approached, and as the disorder in the conductances is decreased \cite{vs98}. 

\section{Conclusion}
We have investigated a dynamic thermal fuse model with heat diffusion. The time dependence is governed by the two competing time scales $\tau_0=gT_rC/i^2$ and $\tau_1=\xi^2/D$. The breakdown time follows $t_r\sim I^{-2}$ in both the small-current domain ($\tau_0>>\tau_1$) and large-current domain ($\tau_0<<\tau_1$). This is in agreement with experiments on electrical breakdown of carbon-polymer composites \cite{Experimental,Predictable2}. In the intermediate domain, competition between the two time scales produces a more complex behaviour. A characteristic feature of this domain is that increasing the thermal diffusivity lengthens the lifetime $t_r$ of the system.  

Heat diffusion introduces new subtleties to the thermal fuse model, which still remain to be investigated. However, the power law behaviour in the divergence of the resistance and in the roughness of the fracture has proven to be robust, and seems independent of  the thermal diffusivity and the disorder in the conductances.

\begin{acknowledgments}
We acknowledge discussions with K. K. Bardhan and Morten Gr\o va. 
This work has been financed by the Norwegian Research Council (NFR) 
Petromax Program No. 174164/S30, StatoilHydro AS, EMGS and Numerical 
Rocks AS. 
\end{acknowledgments}

\bibliography{fm}

\end{document}